\def\lapp{\mathrel{\mathop{<}\limits_{{}^\sim}}}

\documentstyle[12pt]{article}

\title{Parity--violating electron scattering from the pion--correlated
relativistic Fermi gas\thanks{This work is supported in part by
funds provided by the U.S. Department of Energy (D.O.E.) under contract
\#DE-AC02-76ER03069.}}
\author{M.B. Barbaro$^a$, A. De Pace$^a$,
T.W. Donnelly$^b$ and A. Molinari$^{a,c}$
\\ \\
\em $^a$ Dipartimento di Fisica Teorica dell'Universit\`a
 di Torino and \\
\em Istituto Nazionale di Fisica Nucleare, Sezione di Torino, \\
\em via P.Giuria 1, I-10125 Torino, Italy \\
\em $^b$ Center for Theoretical Physics, \\
\em Laboratory for Nuclear Science and Department of Physics, \\
\em Massachusetts Institute of Technology, Cambridge, MA 02139, USA \\
\em $^c$ Ministero degli Affari Esteri -- Consolato Generale d'Italia, \\
\em Boston, MA 02116, USA}

\date{}

\begin{document}

\maketitle

\begin{abstract}

Parity--violating quasielastic electron scattering is studied within
the context of the relativistic Fermi gas and its extensions to include
the effects of pionic correlations and meson--exchange currents. The
work builds on previous studies using the same model; here the part of the
parity--violating asymmetry that contains axial--vector hadronic currents
is developed in detail using those previous studies and a link is
provided to the transverse vector--isovector response. Various
integrated observables are constructed from the differential asymmetry.
These include an asymmetry averaged over the quasielastic peak, as well
as the difference of the asymmetry integrated to the left and right of
the peak --- the latter is shown to be optimal for bringing out the
nature of the pionic correlations.  Special weighted integrals involving the
differential asymmetry and electromagnetic cross section,
based on the concepts of $y$--scaling and
sum rules, are constructed and shown to be suited to studies of the
single--nucleon form factor content in the problem, in particular, to
determinations of the isovector/axial--vector and electric
strangeness form factors.  Comparisons are also made with recent predictions
made on the basis of relativistic mean--field theory.

\end{abstract}

\vfill

CTP\#2223 \hfill July 1993

\eject

\section{Introduction}
\label{sec:intro}

In this paper we investigate the theory behind the inclusive scattering of
longitudinally polarized electrons from nuclei in the kinematical regime
of the quasielastic peak (QEP). Our specific purpose is that of studying
the parity--violating (PV) effects arising from the
interference between the neutral weak and electromagnetic currents which
can be explored through measurement of the helicity asymmetry.
This is a theme that has recently received considerable attention
theoretically [1-8]
and
promises to be an active theme experimentally as well (see, for example,
ref.~\cite{Mus93a} for a review covering past, present and future experimental
prospectives, together with discussions of the connected theoretical
issues).

Notwithstanding the fact that such PV experiments are difficult to carry
out, requiring in particular high luminosity, high electron
polarization, frequently specialized detectors and great care in
controlling systematic errors, the results obtained heretofore
are quite promising and lead one to expect that the new generation of
studies planned at MIT/Bates, Mainz and CEBAF
will bring about significant improvements in our knowledge of subtle
and new aspects of nuclear and nucleonic physics.
In addition, given sufficiently fine information on such hadronic
physics issues, one might even expect that accurate tests
of the standard model in an energy domain far away from the one
explored with high--energy accelerators and yet complementary to atomic
PV studies \cite{Mus93} will be achieved.

In the present work we draw upon some of our past studies of
quasielastic electron scattering \cite{Don92a,Alb88,Alb90},
especially on recent work \cite{Alb93} in which the electromagnetic (EM)
and weak neutral current (NC) {\em vector} responses were explored in
detail. The current study extends that work now to include in--depth
discussion of the {\em axial--vector} response and combines the results
of all of our nuclear physics modeling to yield predictions for the
PV asymmetry.

We focus our attention on the QEP energy region not only because
here the cross sections are large enough to yield significant
figures--of--merit and hence measurable
asymmetries, as discussed in ref.~\cite{Don92a},
but especially because this region is a (perhaps
unique) one in which it should prove possible to unfold
the two facets of the reaction that we wish to explore, namely
those involving nuclear and nucleonic structure and dynamics. For it
is in the QEP region that the process is essentially ``quasi--free'',
implying that the nuclear structure effects are not overwhelming and
suggesting that studies of nucleonic physics (axial--vector and possibly
strangeness form factors of the nucleon, in particular) can be
undertaken. Our study spans a range of momentum transfers extending roughly
from 300 MeV/c up to 1 GeV/c.
Below 300 MeV/c we do not trust the relativistic Fermi gas (RFG)
as a reliable model for
describing nuclear phenomena; for momenta larger than 1 GeV/c severe
difficulties are met in fulfilling Lorentz and gauge invariance in the
extended model. In addition, here at large momentum and energy transfers
$\Delta$ and N$^*$ production become hard to distinguish from quasielastic
scattering and the associated reaction mechanism is not yet sufficiently
under control to allow for a reliable extraction of the single--nucleon form
factors.

For the inputs needed to calculate the PV asymmetry, namely the
electromagnetic and the weak neutral current nuclear responses, we draw, as
already mentioned, on our past work \cite{Alb93}.
Indeed in ref.~\cite{Don92a} the nuclear responses have been calculated
within the framework of the covariant RFG model ---
of course, this model, while representing a good starting point
for nuclear structure studies, needs to be improved upon if it is to be
brought into closer touch with physical reality. With this in mind, the
basic model was extended in ref.~\cite{Alb93} by adding to the RFG the
pion in its dual role of force and current carrier through the
incorporation of selected classes of perturbative diagrams, always taking
great care to respect gauge invariance.
While we believe that it is important to pursue the idea of employing
baryons and mesons as effective degrees of freedom in describing
nuclear phenomena, at least in some limited, intermediate--energy
kinematical domain
(in fact, an {\em ab initio}
treatment in terms of more fundamental degrees of freedom,
namely quarks and gluons, while desirable is as yet beyond reach in this
strong--coupling regime), the restriction of
considering the pion alone among the mesons is certainly questionable.
Yet we also believe, and in this respect we have, hopefully
convincingly, argued in ref.~\cite{Alb93}, that the pion
at least plays an important role in the QEP region and can
be consistently treated at the level of both currents and forces.
Importantly its impact on the nuclear responses is usually only strongly
felt for not too large momentum transfers, defining the critical value of
momentum transfer beyond which such effects become only modest corrections
to the RFG predictions, at least for some of the responses. Not all
observables behave this way, however, and as discussed in ref.~\cite{Alb93}
the PV longitudinal response
in particular, is dramatically modified by pionic isospin--dependent
correlations, producing detectable consequences for
the asymmetry and thus yielding a potentially useful window
on nuclear dynamics. This circumstance is exceptional and when it comes
to addressing instead the physics of the nucleon in
the nuclear medium, our previous studies indicate that the pionic
correlations are sufficiently well understood to
allow us to select with
good confidence the kinematical domains where nucleonic physics inside
nuclei can be safely addressed; that is to say, it is possible to find
situations where the contamination
arising from nuclear structure effects is relatively small, at least as
far as this can be ascribed to pions.

In line with the above considerations, the plan of this paper is the
following. In sect.~2 we summarize the basic formalism involved in PV
electron scattering and begin our discussion of
the axial--vector PV response
(which was not dealt with in ref.~\cite{Alb93}), exploiting
for this purpose its close
relationship with the isovector transverse EM
response in leading order in the non--relativistic reduction.
In sect.~3 we address the question of the impact of pionic correlations on
the inclusive scattering of polarized electrons from nuclei.
This issue appears to be explored best through certain frequency integrals
of the asymmetry which are introduced in sects.~3 and 4 as they are needed.
A particularly important one of these discussed in sect.~3 has the range of
integration divided into two parts, to the left-- and right--hand sides of
the QEP: the right--hand integral is then subtracted from the left--hand
integral to form what is called $\Delta{\cal A}$.
As we shall see, this observable has the property of
emphasizing the role of the (pionic)
correlations in nuclear matter while minimizing the effects of the
single--nucleon form factors.  In sect.~4 several other integrated
quantities are introduced. First, in sect.~4.1 the energy--averaged
asymmetry ${\overline{\cal A}}$ is considered --- this has the merit of
suppressing the
pionic correlation effects while bringing out the dependences on the
single--nucleon form factors, importantly, on the axial--vector form
factor.  However, the degree to which this is accomplished can be
improved upon and, to this end, in sect.~4.2 we introduce two
additional quantities called ${\cal R}_1$ and
${\cal R}_2$
based on previous treatments of scaling \cite{Alb88,Day90} and
sum rules \cite{Alb88,Don89a}. As we shall see in sect.~4.2, these
observables will permit us to extract information on the single--nucleon
axial--vector form factor largely without uncertainties from nuclear
correlation effects and from meson--exchange current (MEC) corrections.
Indeed,
in quantifying the uncertainties from the former we consider not only
the pionic correlation effects discussed in ref.~\cite{Alb93} but also
those treated by Horowitz {\em et al.\/} in ref.~\cite{Hor93} and find
that at large
scattering angles where the axial--vector response has its greatest influence
on the asymmetry in both cases the nuclear--physics--based uncertainties
appear to be well under control.  Finally, in sect.~4 we also consider
the roles played by electric and magnetic single--nucleon strangeness
form factors and then end with our conclusions in sect.~5.

\section{Parity--violating electron scattering}
\label{sec:pvee}

The observable that plays the central role in PV electron scattering
is the asymmetry, defined as
\begin{equation}
  {\cal A} =
  {d^2\sigma^+ - d^2\sigma^-\over d^2\sigma^+ + d^2 \sigma^-}\ ,
\label{eq:asy}
\end{equation}
where $d^2\sigma^+$ and $d^2\sigma^-$ indicate the nuclear
double--differential cross sections for the scattering of
right-- and left--handed
longitudinally polarized electrons.
The asymmetry can be cast in the following
form \cite{Don88,Don89,Don92a,Mus92,Mus93a}
\begin{equation}
  {\cal A} = {\cal A}_0
    {v_L R^L_{AV} (q,\omega) + v_T R^T_{AV} (q,\omega) +
     v_{T'} R^{T'}_{VA} (q,\omega) \over
     v_L R^L(q,\omega) + v_T R^T(q,\omega) }\ ,
\label{eq:asymmetry}
\end{equation}
where
\begin{equation}
  {\cal A}_0 = {G\left| Q^2\right|\over 2\pi\alpha\sqrt{2}}\ ,
\end{equation}
having introduced the lepton kinematical factors \cite{Don86,Ras89}
\begin{eqnarray}
  v_L &=& \left( {Q^2\over q^2}\right)^2 \nonumber\\
  v_T &=& {1\over 2} \left| {Q^2\over q^2}\right|
   + \tan^2 {\theta\over 2} \\
  v_{T'} &=& \sqrt{ \left|{Q^2\over q^2}\right|
   + \tan^2 {\theta\over 2}} \tan {\theta\over2}\ \ .\nonumber
\end{eqnarray}
In the above formulae $q=|{\rm\bf q}|$ is the transferred momentum, $\omega$
the transferred energy, $Q^\mu=(\omega,{\bf q})$, $\alpha$ the fine--structure
constant, $G$ the Fermi constant and $\theta$ the scattering angle.
In (\ref{eq:asymmetry}) $R^L$ and $R^T$ are the usual (purely vector) EM
longitudinal and transverse response functions, respectively. The responses
$R^L_{AV}$, $R^T_{AV}$ and $R^{T'}_{VA}$ all involve interferences between
the EM and NC hadronic matrix elements; the two labeled $AV$ arise from
leptonic axial--vector/hadronic vector contributions, whereas the one labeled
$VA$ arises from the reverse; the two labeled $L$ and $T$ contain only
purely vector hadronic currents (longitudinal and transverse projections,
respectively), whereas the one labeled $T'$ arises from an interference
between the vector EM current and the axial--vector part of the NC and is
purely transverse, although of a different class than the responses
labeled $T$ (see refs.~\cite{Don86,Ras89}). For brevity we shall refer
to these last three PV responses as PV longitudinal, PV transverse and
axial--vector, respectively.

Before exploring how the asymmetry is affected by nuclear correlations
(or by nucleonic form factors) we briefly comment on the level of
accuracy that one can expect to attain in measuring the asymmetry itself.
We start by noting that the statistical precision that one can achieve in
determining ${\cal A}$ with a PV helicity--difference
experiment is given by
\begin{equation}
  \left|{\delta {\cal A}_{\rm expt}\over {\cal A}_{\rm expt}}\right| =
   \left[p_e^2\,{\cal L}\,T\, { { \left( {\overline{{\cal A}\sigma}}
   \right)^2 }
   \over{\overline\sigma} } \right]^{-{1\over2}},
\label{eq:error}
\end{equation}
where $p_e$ is the (longitudinal) polarization of the incident
electron, ${\cal L}$ the luminosity and $T$ the runtime.
Here ${\overline\sigma}$ and ${\overline{{\cal A}\sigma}}$ are integrals of
the EM cross section and that cross section times the asymmetry,
respectively, over the angular and energy
acceptances in the experiment, as well as over any range in $\omega$ that
we wish to consider (see below). We shall be discussing the energy--averaged
asymmetry ${\overline{\cal A}}$ in sect.~4.1 and, to the extent that
${\overline{{\cal A}\sigma}}\approx {\overline{\cal A}}\times
{\overline\sigma}$, one obtains
\begin{equation}
  \left| {\delta {\cal A}_{\rm expt}\over {\cal A}_{\rm expt}}\right| =
   \left[ p_e^2\,{\cal L}\,T\, {\overline{\cal F}} \right]^{-{1\over2}},
\label{eq:errora}
\end{equation}
involving the average figure--of--merit ${\overline{\cal F}}\equiv
{\overline\sigma}\times {\overline{\cal A}}^2$.
In ref.~\cite{Don92a} we have shown that by choosing for these quantities
values that are perhaps presently a little
optimistic, but likely to be attained in the
not--too--distant future, and
by extending the integrations over $\omega$ to include the whole QEP, a
precision of even 1\% can be contemplated for not too large $q$.

An issue that must be dealt with in studying PV quasielastic
electron scattering is the question
of how properly to select the range of the integrations above.
One must generally avoid the
very low and the very high energy tails of the quasielastic spectrum in order,
on the one hand, not to have to deal with specific inelastic nuclear
excitations for which a
description using the RFG is ill--suited, while on the other, not to become
entangled with pion production.
These requirements (in fact the first one is not so serious when
an energy integration is involved) are clearly met by not considering
momentum transfers that are too small or too large. It is also clear that
an integration encompassing most
of the QEP leads to the best statistical precision for ${\cal A}$, although,
as we shall see in the discussions to follow,
in future studies it might become desirable to explore in some detail
the $\omega$--dependence of the asymmetry, accepting the unavoidable
sacrifice in statistical precision that this entails.

On the one hand, in sect.~3 we shall investigate the roles played by pionic
correlations and MEC contributions and so it is useful to recall \cite{Alb93}
that, while the pionic MEC generally play a minor role in the QEP
region, the
correlations brought in by the pion are dominated by
the so--called exchange term, with a characteristic oscillatory
behaviour as a function of $\omega$. Furthermore,
the self--energy contribution, although of minor importance, also
displays the same behaviour. As a consequence, if the
focus is on the correlations, then it appears that the appropriate
procedure is to split the energy integral in two parts --- from the
low--energy end--point to $\omega=\omega_{QEP}\equiv |Q^2|/2m_N$ and
from there to the high--energy end--point.
Then, by subtracting the two contributions, it is
clear that an observable is obtained which maximally emphasizes the impact
of the correlations on the asymmetry. This yields the quantity
$\Delta{\cal A}$ discussed in sect.~3.  It is possible to show that
the precision to be expected for $\Delta{\cal A}$ is much less than for
${\overline{\cal A}}$ (or for the other observables ${\cal R}_{1,2}$ employed
in sect.~4.2 which could be determined to the same precision as
${\cal A}_{\rm expt}$ or ${\overline{\cal A}}$):
\begin{equation}
  \left| { {\delta\ \Delta{\cal A}}\over{\Delta{\cal A}} }\right| \cong
  \left| { {2{\overline{\cal A}}}\over{\Delta{\cal A}} }\right| \times
  \left| { {\delta{\overline{\cal A}}}\over{\overline{\cal A}} }\right| \ .
\end{equation}
Since $|2{\overline{\cal A}}/ \Delta{\cal A}|$ ranges from about 1--7
in the forward direction and 10--20 in the backward direction
as $q$ goes from 300 MeV/c to 1 GeV/c, the fractional precision in
$\Delta{\cal A}$ will typically be around a few \% in the forward
direction and 10--20\% in the backward direction at best.  Although this
is much poorer than for the integrated quantities discussed below, it
nevertheless will be shown in sect.~3 to be good enough to provide an
interesting window on the pionic physics issues.

On the other hand, in sect.~4 we shall address the issue of the sensitivity
in the PV asymmetry to variations in the isovector/axial--vector and
magnetic and electric strangeness form factors of the nucleon (see
sect.~2.1).  For this purpose we wish to suppress the sensitivity to
pionic correlation and MEC effects in order to reveal the dependences on
the form factors. Motivated by concepts of $y$--scaling and electroweak
sum rules where these effects can be de--emphasized, accordingly we shall
perform various integrals involving the differential asymmetry and the EM
cross section (usually with specific weighting factors) to define new
observables, ${\overline{\cal A}}$ in sect.~4.1 and ${\cal R}_{1,2}$ in
sect.~4.2.  There the integrations will extend over the entire region
of the quasielastic response, even though, in the future in the context of
a particular experiment, it will probably be necessary to address the
problem of pion production in the high--$\omega$ part of this region.

\subsection{Single--nucleon form factors}
\label{subsec:formfactors}

As has already been mentioned above, the responses
appearing in (\ref{eq:asymmetry}) have been calculated in a companion
paper \cite{Alb93} for a pion--correlated RFG, except for the axial--vector
response $R^{T'}_{VA}$ which will be addressed in the next subsection.
For this purpose, we start by recalling that the nucleonic
electromagnetic and weak neutral vector ($V$) and axial--vector
($A$) currents read
\begin{eqnarray}
  J^\mu_{EM}(Q,P) &=&
   \bar{u}(P+Q,s')\left[F_1(Q^2)\gamma^\mu
    +i{F_2(Q^2)\over2m_N}\sigma^{\mu\nu}Q_\nu\right]u(P,s) \\
  J^\mu_{NC,V}(Q,P) &=&
   \bar{u}(P+Q,s')\left[{\widetilde F}_1(Q^2)\gamma^\mu
    +i{{\widetilde F}_2(Q^2)\over2m_N}\sigma^{\mu\nu}Q_\nu\right]u(P,s) \\
  J^\mu_{NC,A}(Q,P) &=& \bar{u}(P+Q,s')\,\widetilde{G}_A(Q^2)\,
   \gamma^5\gamma^\mu u(P,s)\ ,
\end{eqnarray}
where $s$ and $s'$ are the third components of the nucleon spin,
$F_1$ and $F_2$ the standard Dirac and Pauli nucleon form factors (and
${\widetilde F}_{1,2}$ their vector NC analogs),
$\widetilde{G}_A$ the axial--vector NC form factor and
$P^{\mu}=(E({\bf p}),{\bf p})$
the on--shell nucleon's four--momentum in the Fermi sphere.
In fact, it is common practice in EM studies
of the nucleus to use the Sachs \cite{Sac62} electric and
magnetic form factors $G_E$ and $G_M$ rather than $F_1$ and $F_2$:
$G_E=F_1-\tau F_2$ and $G_M=F_1+F_2$, with $\tau\equiv |Q^2|/4m_N^2$.
Analogous expressions can, of course, be written for NC Sachs--like
form factors ${\widetilde G}_E$ and ${\widetilde G}_M$. All form
factors in addition can be labeled with $p$ or $n$ to denote which
are for the proton and which for the neutron, respectively, or with
$T=0,1$ to denote which are isoscalar or isovector, respectively.

For the convenience of the reader and for sake of completeness
we briefly summarize the parameterization of the
single--nucleon form factors employed in the present research.
For a broader discussion of the subject we refer the
reader to refs.~\cite{Mus92,Mus93a}.
For the electric and magnetic nucleonic EM form factors we have used the
standard forms
\begin{eqnarray}
  G_{Ep}(\tau) &=& G_D^V (\tau) \\
  G_{Mp} (\tau) &=& \mu_p G_D^V (\tau) \\
  G_{Mn} (\tau) &=& \mu_n G_D^V (\tau) \\
  G_{En} (\tau) &=& - \mu_n \tau G_D^V (\tau) \xi_n (\tau)\ \ ,
\label{eq:GEn}
\end{eqnarray}
where $G_D^V(\tau) = \left[ 1 + \lambda_D^V \tau\right]^{-2}$ is the
vector dipole form factor with $\lambda_D^V \cong 4.97$ and where
$\mu_p\cong 2.793$ and $\mu_n \cong -1.913$ are the proton and neutron
magnetic moments, respectively.
Formula (\ref{eq:GEn}), with
$\xi_n(\tau) = \left[ 1 + \lambda_n \tau\right]^{-1}$
and $\lambda_n = 5.6$, is usually referred to as the Galster
parameterization \cite{Gal71}.
For the isovector weak axial--vector form factor we use
\begin{equation}
  G^{(1)}_A (\tau) = g^{(1)}_A G_D^A (\tau)\ \ ,
\label{eq:GAvec}
\end{equation}
where $G_D^A(\tau) = \left[ 1 + \lambda_D^A\tau\right]^{-2}$ is the
axial--vector dipole form factor, $g^{(1)}_A=1.26$ from neutron
$\beta$--decay and we have taken $\lambda_D^A\cong 3.53$.  The NC
(strong isospin 1 channel $=$ isovector) form factors are then
\begin{eqnarray}
  {\widetilde G}^{(1)}_E(\tau) &=& \beta_V^{(1)} G_E^{(1)} \\
  {\widetilde G}^{(1)}_M(\tau) &=& \beta_V^{(1)} G_M^{(1)} \\
  {\widetilde G}^{(1)}_A(\tau) &=& \beta_A^{(1)} G_A^{(1)}  \ .
\end{eqnarray}
In the standard model the tree--level electroweak hadronic couplings
are given by $\beta_V^{(1)}=1-2\sin^2 \theta_W$,
$\beta_A^{(1)}=1$ (isovector), and for
use below, $\beta_V^{(0)}=-2\sin^2 \theta_W$,
$\beta_A^{(0)}=0$ (isoscalar), using the notation of ref.~\cite{Don79}.
Here $\theta_W$ is the weak mixing angle ($\sin^2 \theta_W\cong 0.227$).

Since we shall also be concerned with the
strangeness content of the nucleon,
we have introduced in the NC (strong isospin 0 channel $=$ isoscalar) form
factors
\begin{eqnarray}
  {\widetilde G}^{(0)}_E(\tau) &=& \beta_V^{(0)} G_E^{(0)}-G_E^{(s)} \\
  {\widetilde G}^{(0)}_M(\tau) &=& \beta_V^{(0)} G_M^{(0)}-G_M^{(s)} \\
  {\widetilde G}^{(0)}_A(\tau) &=& \beta_A^{(0)} G_A^{(0)}-G_A^{(s)}  \ ,
\label{eq:GAz}
\end{eqnarray}
the following purely isoscalar strangeness form factors:
\begin{eqnarray}
  G^{(s)}_E(\tau) &=& \rho_s \tau G_D^V(\tau) \xi^{(s)}_E(\tau)
    \label{eq:GEs} \\
  G^{(s)}_M(\tau) &=& \mu_s G_D^V(\tau)
    \label{eq:GMs} \\
  G^{(s)}_A(\tau) &=& g^{(s)}_A G_D^A(\tau) \ ,
\end{eqnarray}
where again $G_D^V(\tau)$ and $G_D^A(\tau)$ are the dipole form factors
used above and where $\xi^{(s)}_{E} \equiv \left[1 + \lambda^{(s)}_{E}
\tau\right]^{-1}$.
As an orientation, we shall let the strength of the electric and
magnetic strangeness form factors vary in the ranges \cite{Mus92,Mus93a}
\begin{equation}
  \rho_s \colon 0 \to - 3 \qquad \mu_s \colon 0 \to - 1\ \ ,
\end{equation}
whereas for $\lambda^{(s)}_E$ we shall consider the two options
$\lambda^{(s)}_E=0$ and $\lambda^{(s)}_E=\lambda_n=5.6\,$.
Finally, in our analysis we shall set $g^{(s)}_A=0$, since it plays only
a minor role in quasielastic electron scattering.  Note that these
expressions have been stated at the tree level; beyond tree level we may
allow for additional contributions in the form $g_A^{(1)}\rightarrow
g_A^{(1)}[1+R_A^{(1)}]$, {\em etc.\/} to take into account radiative
corrections (this is the approach followed in refs.~\cite{Mus92,Mus93a})
or we may consider the above choices for the parameters $\{ g_A^{(1)}\dots
\}$ to be starting approximations in fixing our parameterizations of
the effective form factors that arise
when higher--order electroweak corrections are taken into account.  Of
course, the effective form factors then become process--dependent.
Specifically,
from modeling of such contributions \cite{Mus90}, one expects that the
effective $g_A^{(1)}$ could differ from 1.26 by as much as $\pm$20\%.
Indeed, putting the burden on the strength of the form factors is not
the most general way of effectivizing the problem: for example, the
$|Q^2|$--dependences in many cases are not known or are only known with
limited precision.  However, our interest in the present work when
any aspect of the single--nucleon form factors is concerned is to
explore whether or not variations ranging from a few \% to as much as
10\% away from the starting parameterizations would be manifest in
specific observables.  Whether those variations occur because of
higher--order electroweak contributions or are due to differences in
the $|Q^2|$--dependences, the relevant issue to address here is the
level of sensitivity in quasielastic PV electron scattering to such
variations.

\subsection{The axial--vector response}
\label{subsec:axial}

Following refs.~\cite{Don92a,Alb93,Alb88,Alb90},
the above currents are to be inserted into the so--called hadronic
interference tensor for the RFG
\begin{eqnarray}
  W^{\mu\nu}_{EM/NC} &=&
   {3{\cal N} m^2_N\over 4\pi p^3_F} \int {d^3p\over
   E({\bf p}) E({\bf p}+{\bf q})}
   \delta\left\{\omega-\left[E({\bf p}+{\bf q})-E({\bf p})\right]
   \right\} \nonumber\\
  &&\times\theta\left(p_F-|{\bf p}|\right)
   \theta\left(|{\bf p}+{\bf q}| -
   p_F\right) f^{\mu\nu}_{EM/NC} (P+Q,P)\ ,
\end{eqnarray}
where $p_F$ is the Fermi momentum,
$E({\bf p}) = \left[ {\bf p}^2 + m^2_N\right]^{1/2}$ the energy of a nucleon
with momentum ${\bf p}$, ${\cal N}$ the particle number
($Z$ for a proton gas, $N$ for a neutron one) and
\begin{equation}
  f^{\mu\nu}_{EM/NC}(P+Q,P) = {1\over2}{\rm Tr}
  \left[J^{\mu}_{EM}(Q,P)J^{\nu\dagger}_{NC,A}(Q,P) +
  J^\mu_{NC,A}(Q,P)J^{\nu\dagger}_{EM}(Q,P)\right] \ ,
\label{eq:femnc}
\end{equation}
the single--nucleon interference tensor ({\em i.e.,\/} only the part
which contains axial--vector hadronic currents; the corresponding
vector contributions have been discussed in ref.~\cite{Alb93}). Then
in symmetric nuclear matter (${\cal N}=Z=N=A/2$) one obtains the
axial--vector, transverse, purely
isovector\footnote{Actually the standard model
beyond the tree level allows for an isoscalar component of the axial--vector
current of the nucleon (see (\ref{eq:GAz}) above),
which we shall disregard in the present study.} PV
nuclear response:
\begin{eqnarray}
  R^{T'}_{VA} &\equiv& -i a_V W^{12}_{EM/NC} \nonumber\\
  &=& a_V {3{\cal N}\over 4m_N \kappa \eta^3_F} (\epsilon_F
  - \Gamma) \theta(\epsilon_F - \Gamma)
  \sqrt{\tau(1+\tau)}\, G_M^{(1)}(\tau) G_A^{(1)}(\tau)
  \left\{ 1 + \widetilde\Delta\right\} ,
\label{eq:RTprime}
\end{eqnarray}
$a_V=-(1-4\sin^2\theta_W)$ being the standard model leptonic vector
coupling constant (for completeness, $a_A=-1$ is the standard model
leptonic axial--vector coupling constant).
In (\ref{eq:RTprime}) we employ the functions \cite{Don92a,Alb88}
\begin{equation}
  \Gamma(q,\omega)\equiv
    {\rm max} \left\{ \left( \epsilon_F - 2\lambda\right),\
    \kappa\sqrt{1+{1\over \tau}} -\lambda\right\}
\end{equation}
and
\begin{equation}
  \widetilde\Delta \equiv
   {1\over\kappa} \sqrt{\tau\over 1+\tau}\left\{ {1\over 2}
   (\epsilon_F +\Gamma) + \lambda\right\} - 1 \ \ ,
\label{eq:Cdelta}
\end{equation}
having introduced the dimensionless variables:
\begin{eqnarray}
  &&\left.
   \begin{array}{rcl}
   \kappa  &\equiv& {q/ 2m_N} \\
   \lambda &\equiv& {\omega/ 2m_N}
   \end{array} \right\}
   \Longrightarrow \tau=\kappa^2 -\lambda^2 \nonumber\\
   \\
    &&
   \!\!\begin{array}{rcl}
   \eta_F &\equiv& {p_F/ m_N} \\
   \varepsilon_F &\equiv& {E({\bf k}_F)/ m_N} =
    \sqrt{ 1 + \eta^2_F}\ \ .
   \end{array}  \nonumber
\end{eqnarray}

Next we consider the leading order of the non--relativistic expansion
of the space components of the electromagnetic and weak axial--vector currents.
They read
\begin{eqnarray}
   {\bf J}_{EM} &\approx& -i G_M\chi^\dagger_{s'}
    {(\mbox{\boldmath $\sigma$}\times{\bf q})\over2m_N}\chi_s
  \label{eq:Jem} \\
  {\bf J}_{NC,A}  &\approx&
   -\widetilde{G}_A\chi^\dagger_{s'}\mbox{\boldmath $\sigma$}\chi_s\ ,
  \label{eq:JA}
\end{eqnarray}
where $\chi_s$ is a 2--component (spin) spinor.
By inserting these expressions in (\ref{eq:femnc}), one obtains for
the space components of the single--nucleon interference hadronic
tensor
\begin{equation}
  f^{ij}_{EM/NC} = {i\over m_N} G_M \widetilde{G}_A \epsilon_{ijk}q_k\ .
\end{equation}
On the same footing, by calculating $f^{ij}_{EM}$ with (\ref{eq:Jem}),
one gets
\begin{equation}
  f^{ij}_{EM} = {1\over4m_N^2} G_M^2 (q^2\delta_{ij}-q_i q_j)\ ,
\end{equation}
which would yield, when embedded in the hadronic EM
tensor, the non--relativistic expression for the transverse
EM response according to
\begin{equation}
  R^T=W^{11}_{EM}+W^{22}_{EM}\ .
\end{equation}
Notably, it then turns out that the isovector component of the
latter, $R^{T(1)}$, and the PV axial--vector response in
the leading order of the non--relativistic expansion
(obtained via (\ref{eq:Jem}) and (\ref{eq:JA})) are
{\em exactly} connected through
the simple formula
\begin{equation}
  R^{T'}_{VA}(q,\omega) = a_V {G_A^{(1)}\over G_M^{(1)}}{1\over\kappa}
  R^{T(1)}(q,\omega)\ .
\label{eq:RTprimenonrel}
\end{equation}
Actually, by comparing the exact expressions for $R^{T(1)}$ and
$R^{T'}_{VA}$ (see ref.~\cite{Don92a}), the link (\ref{eq:RTprimenonrel})
can be extended into the relativistic regime.
Indeed, it turns out that the prescription
\begin{equation}
  R^{T'}_{VA}(q,\omega) \cong a_V
   {G_A^{(1)}\over G_M^{(1)}}\sqrt{\tau+1\over\tau}
   R^{T(1)}(q,\omega)
\label{eq:RTprimeapp}
\end{equation}
holds to better than 2\% in the momentum range 300 MeV/c $<q<1$ GeV/c.

In Fig.~1 we show the response $R^{T'}_{VA}$ for the free RFG (dashed
curve) and for the model with pionic correlations and MEC effects included
(solid curve) and so
complete the picture of the five electroweak responses that we began in
ref.~\cite{Alb93}.  We see a ``hardening'' (a shift to higher $\omega$)
of the peak of the response at intermediate values of $q$ which then
fades away and even leads to a slight ``softening'' of the response at
the highest momentum transfers considered.
In computing the response with pionic effects
included we have used (\ref{eq:RTprimeapp}) to provide a link between
the response $R^{T(1)}$ obtained previously, even though that relationship
was derived for the free RFG.
When one introduces such effects in $R^{T'}_{VA}$
the question arises whether this axial--vector/transverse EM link
holds as well in this instance.
 In this connection an important simplifying result
 holds: by performing the appropriate spin algebra,
 one easily sees that in the leading order of the non--relativistic
 expansion for {\em any} Feynman diagram at whatever order one
 always has (schematically)
 $R^{T'}_{VA}\sim\sigma_1\;(\mbox{\boldmath $\sigma$}\times{\bf q})_2 +
                 (\mbox{\boldmath $\sigma$}\times{\bf q})_1\;\sigma_2
             \sim\sigma_1\;\sigma_1 + \sigma_2\;\sigma_2 $
 and
 $R^T\sim(\mbox{\boldmath $\sigma$}\times{\bf q})_1 \;
         (\mbox{\boldmath $\sigma$}\times{\bf q})_1 +
         (\mbox{\boldmath $\sigma$}\times{\bf q})_2 \;
         (\mbox{\boldmath $\sigma$}\times{\bf q})_2
     \sim\sigma_1\;\sigma_1 + \sigma_2\;\sigma_2 $,
 so that (\ref{eq:RTprimenonrel}) is an exact relation to all orders.
 Since the prescription given in our previous work to get approximate
 relativistic response functions from their non--relativistic counterparts
 does not involve their spin structure, the validity of
 (\ref{eq:RTprimeapp}) in the relativistic regime is then inferred.
The door is thus open for the calculation of the nuclear axial--vector
response to the level of accuracy of (\ref{eq:RTprimeapp}), since
the necessary ingredients, namely the various contributions to the
electromagnetic isovector transverse nuclear
response, have been calculated in the work reported in ref.~\cite{Alb93}.
One caveat should, however, be added to the above considerations:
because of the isovector, transverse nature of the axial--vector
response, the MEC contributions of the type shown in Fig.~2, when the
axial--vector part of the $Z^0$ coupling is involved,  are
of higher order in the
non--relativistic expansion (Kubodera--Delorme--Rho
theorem \cite{XXXDKR}) and have accordingly been disregarded.  At high
enough energy/momentum in future work it may be necessary to re-examine this
approximation in more depth.  Of course, as discussed in ref.~\cite{Alb93},
when the vector part of the $Z^0$ coupling is involved such MEC effects
are taken into account in the present work.

\section{Pionic correlations in the asymmetry}
\label{sec:sigmaminus}

In accordance with the arguments given in the previous section here we discuss
the observable
\begin{equation}
  \Delta{\cal A}(q,\theta) \equiv {1 \over{\Delta\omega}}\Bigl[
  \int_{\omega_{min}}^{\omega_{QEP}}\!\!d\omega
   {\cal A}(\theta;q,\omega) -
  \int_{\omega_{QEP}}^{\omega_{max}}\!\!d\omega
   {\cal A}(\theta;q,\omega)\Bigr]\ ,
\label{eq:sigmin}
\end{equation}
where $\omega_{min}$ and $\omega_{max}$ are the RFG response boundaries
for a fixed $q$,
\begin{eqnarray}
  \omega_{min} &=& \sqrt{(p_F-q)^2+m_N^2}-\sqrt{p_F^2+m_N^2}
    \label{eq:omegamin}\\
  \omega_{max} &=& \sqrt{(p_F+q)^2+m_N^2}-\sqrt{p_F^2+m_N^2}\ ,
    \label{eq:omegamax}
\end{eqnarray}
and $\omega_{QEP}=|Q^2|/2m_N$, as above.  We shall make use of the
energy interval
\begin{eqnarray}
  \Delta\omega &\equiv& \omega_{max} - \omega_{min}
    \label{eq:domega}\\
  &=& \sqrt{(p_F+q)^2+m_N^2} - \sqrt{(p_F-q)^2+m_N^2}
    \label{eq:domegb}\\
  &\cong& 2 p_F q/\sqrt{q^2+m_N^2}
    \label{eq:domegc}
\end{eqnarray}
and so define it here.
The expressions given here pertain in the case where $q>2p_F$; of course, when
results are presented for $q<2p_F$ (the Pauli--blocked region) the correct
equations are used.

In the present paper we have calculated $\Delta{\cal A}$ taking into
account all
Feynman diagrams with one pion line: these include the self--energy,
the exchange and the MEC contributions, all of which have been extensively
dealt with in our past work. For sake of illustration we have set
the Fermi momentum $p_F=225$ MeV/c, which roughly corresponds to a
light nucleus such as $^{12}$C.

That the pionic correlations are particularly felt by $\Delta{\cal A}$,
as previously anticipated, is clearly
apparent from Fig.~3. Indeed, there we first observe that at $q=300$ MeV/c
in the case of results obtained with the free relativistic Fermi gas
(labeled RFG)
$\Delta{\cal A}_{RFG}$ almost vanishes because of the nearly perfect
cancellation between the contributions where $\omega<\omega_{QEP}$ and
those where $\omega>\omega_{QEP}$;
however, at larger $q$ this cancellation
becomes less complete, owing partly to the role played by the
nucleonic form factors and partly to the RFG model itself, whose responses
(in contrast to the non--relativistic case) become less and less
symmetric as $q$ increases.
It is also clear from Fig.~3 that the correlations, in particular the
exchange contributions,
dramatically alter the prediction of the free RFG, yielding a huge
$\Delta{\cal A}_{\pi}$ at small $\theta$.
This outcome is simply interpreted: of the nuclear responses
that enter in the asymmetry the pion has its greatest effect on
$R^L_{AV}$ (see ref.~\cite{Alb93}) and although the latter
in the RFG model accounts only for at most about 10\% of the total
asymmetry (and this only in the forward direction), nevertheless the impact of
the pionic correlations is violent enough to induce a
large negative value of $R^L_{AV}$ at small
$\omega$, which is in turn reflected in the large negative value
of $\Delta{\cal A}_{\pi}$ at small $\theta$ displayed in Fig.~3.
We deduce from this that the characteristic behaviour of $\Delta{\cal A}$ with
$\theta$ shown in the figure
represents one of the most transparent signatures of
pion--induced isoscalar correlations in nuclei (we recall that $R^{T'}_{VA}$
is purely isovector and that in $R^T_{AV}$ the isoscalar contribution is
strongly suppressed --- see ref.~\cite{Alb93}). We return to the families
of curves in Fig.~3 in the discussions below.

Let us next examine the $\omega$--dependence of
the asymmetry. Looking first at Fig.~4, we notice the significant effect
occurring at moderate $q$
(say 300--500 MeV/c), small $\omega$ and forward angles that is
responsible for the striking behaviour of $\Delta{\cal A}$ previously commented
upon.
As discussed above, it is related to the large negative value assumed by the
correlated $R^L_{AV}$, which leads to a pionic asymmetry that is
an order--of--magnitude larger than the free RFG one.
However, as $\omega$ increases
$R^L_{AV}$ rapidly decreases until it changes sign, while $R^T_{AV}$
stays negative: accordingly, they largely cancel in the numerator
of the ratio expressing ${\cal A}$ and this becomes substantially lowered.

Interestingly, an energy is reached (about 60 MeV for $q=300$ MeV/c)
where the correlated and free RFG values of ${\cal A}$ coincide.
At still larger $\omega$ a further reduction of
${\cal A}$ is seen to occur until, at about 90 MeV, it nearly vanishes.
This constitutes an example of a dynamical {\em restoration} of a
symmetry (here the left--right parity symmetry) and reflects the complex
nature of the PV longitudinal response. Indeed, for the free RFG
\begin{equation}
R^L_{AV} = a_A\left[\beta^{(0)}_V R^{L(0)}_{EM}
  + \beta^{(1)}_V R^{L(1)}_{EM}\right]
\end{equation}
is very small owing to the internal fight between its isoscalar and
isovector components brought about by the opposite signs and similar
magnitudes of the standard model hadronic couplings,
$\beta^{(0)}_V\cong-0.45$ and $\beta^{(1)}_V\cong0.55$, as discussed in
ref.~\cite{Don92a}. In fact, the free RFG response $R^L_{AV}$ is negative
at $q=300$ MeV/c, becomes positive at
500 MeV/c and then stays so up to very large momenta,
always remaining quite small in magnitude.
When pionic correlations are switched on, this behaviour
is profoundly altered.
In particular, the magnitude of $R^L_{AV}$ is
much increased, becoming comparable in magnitude to, for example, the
EM response $R^L$. In addition, for a given moderate value of $q$
the response $R^L_{AV}$ changes sign at some $\omega$.
The near vanishing of the asymmetry at $\omega\approx90$ MeV in
Fig.~4 thus stems from the cancellation between the
{\em positive} contribution it gets from $R^L_{AV}$ and the
{\em negative} one it gets from $R^T_{AV}$.
This trend of course fades away at larger $\theta$, where the role
of the longitudinal PV response gradually becomes irrelevant.
Finally, from Fig.~4, we see that at larger momenta, where the impact of
correlations is no longer so strongly felt, the nearly perfect
restoration of the left--right symmetry does not show up anymore.
It is, however, still true (even
at 1 GeV/c) that an energy exists where the free and the correlated
values of the asymmetry coincide.

We now conclude this section with a brief discussion of
the influence of the nucleonic form factors on the asymmetry.
As previously stated, $\Delta{\cal A}$ has been specifically devised to
enhance the signal for nuclear correlations.
That this is indeed the case can be inferred from Figs.~3 and 4
where we display results
allowing for a variation of the strength $g^{(1)}_A$ of the isovector
axial--vector form factor of $\pm10$\% around the canonical value
$g^{(1)}_A=1.26$ (Cf. the discussion on the uncertainty in the effective
axial--vector form factor in the previous section).
As one can see from Fig.~3, $\Delta{\cal A}$ is totally insensitive to this
variation of
$g^{(1)}_A$ at low momenta; it becomes mildly so at larger momenta,
but even in this case the impact on $\Delta{\cal A}$ of pionic correlations
remains an order--of--magnitude larger than the that arising from
variations in the axial--vector form factor.
Note that an angle $\Theta^-$ exists
($\Theta^-\sim110^o$, relatively independent of $q$)
such that for 500 MeV/c $\lapp q \lapp$ 1 GeV/c
the free and correlated $\Delta{\cal A}$ coincide.
A compensation is thus seen to occur inside $\Delta{\cal A}$ among the
transverse, axial--vector and longitudinal response functions
in the two cases.  In Fig.~4, where the $\omega$--dependence of the
asymmetry is displayed, we again see the influence of variations in the
effective axial--vector form factor: clearly the effects are greatest
at large angles, where $R^{T'}_{VA}$ is weighted most heavily.

Finally, in Figs.~5 and 6 we show results like those above, but now
for variations of the magnetic (Fig.~5) and electric (Fig.~6)
strangeness form factors of the nucleon. We have limited our focus
to $q=$ 500 MeV/c and to rather forward and backward angles, although
the results are representative of other kinematical situations.
Clearly, as expected (see ref.~\cite{Don92a}) the sensitivity to
variations in the amount of magnetic strangeness is very weak (Fig.~5).
Similarly, while not quite as weak, the dependence on electric
strangeness in Fig.~6 as represented by the rather liberal variations employed
here (see the discussions in refs.~\cite{Mus92,Mus93a}) is still
overwhelmed by the effects of correlations.  At higher values of $q$
than those displayed here and for forward--angle scattering, the
sensitivity to variations in $G_E^{(s)}$ grows sufficiently to compete
with the effects of the correlations and consequently the observables
discussed in this section become less well suited to use in attempting
to disentangle the single--nucleon from the many--body effects. In the
next section, we shall discuss observables that are better designed to
accomplish this.

\section{The parity--violating asymmetry: isovector/axial--vector and
strangeness form factors of the nucleon}
\label{sec:sigmaplus}

In this section we introduce and discuss several observables that can
be constructed from the PV asymmetry and the EM cross section which
are designed to minimize the effects of pionic correlations (and, as
we shall see, apparently other types of many--body effects as well) and
hence allow us access to the single--nucleon form factors.
Specifically, when our focus is the isovector/axial--vector
single--nucleon form factor we shall be interested in large scattering
angles where longitudinal effects fade away and where $R^{T'}_{VA}$
has its largest effect (see ref.~\cite{Don92a}). Under these
circumstances the asymmetry becomes
\begin{equation}
  {\cal A}\rightarrow{\cal A}_0{v_T R^T_{AV}(q,\omega) +
    v_{T'}R^{T'}_{VA}(q,\omega)
    \over v_T R^T(q,\omega)}\ .
\label{eq:AT}
\end{equation}
Now, although the isoscalar piece of $R^T$ is generally small, involving
as it does the square of the isoscalar magnetic moment, it is still
important to take it into account when attempting high--precision
determinations of $R^{T'}_{VA}$ (see refs.~\cite{Hor93,Alb93}). To explore
at little further
why this is an issue, let us divide both numerator and denominator of
(\ref{eq:AT}) by $v_T R^{T(1)}$ to obtain
\begin{equation}
  {\cal A} = -{1\over2}{\cal A}_0 \left\{ [1+(1-4\sin^2\theta_W)]
   -{2\over{1+\rho}}(\rho+\rho')\right\}
\label{eq:ATapp}
\end{equation}
for large $\theta$, where we have defined the following ratios of
response functions:
\begin{eqnarray}
  \rho &\equiv& {R^{T(0)}\over R^{T(1)}} \\
  \rho' &\equiv& {R^{T'}_{VA}\over R^{T(1)}} \ .
\end{eqnarray}
The problem is the competition between $\rho'$, which we wish to
determine, and $\rho$ in the second term in (\ref{eq:ATapp})
involving the combination $\rho+\rho'$.  The fractional uncertainty in
the former may be expressed in the following way:
\begin{eqnarray}
  \left| { {\delta\rho'}\over{\rho'} }\right| &\cong&
  \left\{ \left[ {1\over{\rho'}}({{\cal A}
  \over{{\cal A}_0}}) {{\delta{\cal A}}\over{\cal A}}\right]^2 +
  \left[ {\rho\over{\rho'}}{{\delta\rho}\over{\rho}}\right]^2 \right\}^{1/2} \\
  &\cong& \left\{ \left[ 5{{\delta{\cal A}}\over{\cal A}}\right]^2 +
  \left[ 0.5{{\delta\rho}\over{\rho}}\right]^2 \right\}^{1/2}\ ,
\end{eqnarray}
where the last form results from using actual values for the responses
and asymmetry at $q\sim 500$ MeV/c yielding $\rho\sim 0.05$ and
$\rho'\sim -0.1$. An uncertainty of $\sim$1\% in the asymmetry or $\sim$10\%
in the ratio $\rho$ then
produces a corresponding uncertainty of $\sim$5\% in the quantity of interest,
$\rho'$.  Thus we see that several strategies are suggested as ways to
proceed: one is to use parity--conserving (EM) electron scattering to
limit the freedom in the model to the extent that this can be done (see
below); another is to use PC {\em and} PV electron scattering to learn
more about the correlation effects --- this was discussed in the last
section and there we saw considerable sensitivity to the pionic effects which
could serve to make the uncertainty in $\rho$ rather small using
measured values for $\Delta{\cal A}$; a third strategy, the one
adopted in this section, is to form specific weighted integrals
involving the asymmetry and the EM cross section that suppress the
pionic correlation effects embodied in $\rho$ and hence obtain new
observables that are especially suited to determining the single--nucleon
dependences. We begin with the energy--averaged asymmetry.

\subsection{The energy--averaged asymmetry}
\label{subsec:avgasy}

First, we examine the physical observable
\begin{equation}
  {\overline{\cal A}}(q,\theta) \equiv {1\over{\Delta\omega}}
    \int_{\omega_{min}}^{\omega_{max}}\!\!d\omega\,
    {\cal A}(\theta;q,\omega)
\end{equation}
with the limits given in (\ref{eq:omegamin}) and (\ref{eq:omegamax}),
where $\Delta\omega$ is given in (\ref{eq:domega}).  Being an average
over some range in energy, one might hope that the exact values taken for
the end--point energies will not be crucial, although it should be realized
that the energy interval involved
is not dictated by compelling physical arguments, but simply reflects
our theoretical framework which is restricted to particle--hole (ph)
excitations and ignores electroproduction of mesons or
internal excitations of the nucleon.
The excitation of the $\Delta$, for example,
can indeed affect ${\overline{\cal A}}$ via the low--energy tail of its
response
function. As stated above, we shall not allow $q$ to be so large that
consideration of such effects becomes inevitable and thus we
set aside these problems in the present work and leave them for
future research.

As already anticipated above, in contrast to $\Delta{\cal A}$ discussed
above, ${\overline{\cal A}}$ should be rather insensitive to the pionic
correlations as these tend to cancel out in such a symmetrical integral.
The MEC contribution, on the other hand, does not average out in
${\overline{\cal A}}$, although in the $ph$
sector of the nuclear excitations this turns out to be rather puny.
Typical results are shown in Fig.~7 for $q=$ 500 MeV/c.
In particular, as seen in the expanded view in Fig.~7b, at very
backward scattering angles where one might hope to determine the
effective axial--vector coupling $g_A^{(1)}$ (again variations of
$\pm 10$\% around 1.26 are shown in the figure) the free RFG and pionic
correlated results for the energy--averaged asymmetry come together
(for this momentum transfer at $\theta\approx 147^o$).  At other momentum
transfers the angle at which the two models come together is different
from $147^o$ --- since this special condition is presumably model--dependent,
it is unlikely that one can count on using such particular kinematics
to effect a determination of $g_A^{(1)}$ through the variations shown
in the figure.  As we shall see in sect.~4.2, other observables are
better suited for that purpose in any event.

In contrast to the backward--angle situation,
at forward scattering angles where the pionic correlations
induce drastic modifications in $R^L_{AV}$, as we have seen, here the two
families of curves differ, although certainly not as much as in the
case of $\Delta{\cal A}$ discussed in the previous section. In other words,
the observable ${\overline{\cal A}}$ has some of the properties that we are
looking for when we construct quantities that suppress the effects of
correlations while bringing out the dependences on the single--nucleon
form factors; however, this particular observable appears not to be
entirely optimal.  Since the PV longitudinal response is so strongly
affected by the presence of pionic correlations, it is necessary to
adopt an alternative approach to minimize these effects and this is
the subject of the next subsection.

\subsection{The scaling and sum rule approaches}
\label{subsec:sumrule}

In order to free ourselves from the dependence on correlation effects, here
we operate on the differential asymmetry with a more elaborate procedure
than in the previous subsection.
In fact, instead of performing a simple integration over
${\cal A}(\theta;q,\omega)$, we shall now integrate {\em separately}
over the numerator and the denominator of the asymmetry including specific
weighting factors before taking their ratio: in
this way we are naturally led to consider standard scaling relations and
sum rules for the basic nuclear EM and PV responses.
A well--known theorem \cite{Tak90} of many--body theory then insures that the
frequency integrals we are considering are not going to be affected by
self--energy terms and accordingly we should
clearly obtain physical observables that are largely independent of
such effects. These new observables will be seen to be well--suited for
studies of the single--nucleon form factors.

The guiding principle in forming the new observables from the differential
asymmetry and EM cross section is to remove the dependence on the
single--nucleon form factors that varies most rapidly with $\omega$ before
performing the integrals. One possibility is to proceed as in
considerations of $y$--scaling \cite{Alb88,Day90} where a prescription
has been devised in the past that allows one to accomplish this
task, if not exactly (actually, this is impossible in a relativistic context),
at least to a very good level of approximation. Since we already know that the
$R^T_{AV}$ response is dominant in the numerator of the ratio we will be
forming \cite{Don92a} and, moreover, we wish to design that ratio to
work best for backward--angle electron scattering in order to be able to
extract information about the effective coupling $g_A^{(1)}$, we choose
weighting factors that accomplish this ideally for $R^T_{AV}$ (numerator)
and $R^T$ (denominator).

According to the procedure given in ref.~\cite{Alb88}, in order to obtain
quantities that have good $y$--scaling behaviour (called $\psi$--scaling
in that reference for the reasons presented there)
we are to divide the EM and PV responses
\begin{eqnarray}
W^{EM} &=& v_L R^L +v_T R^T \\
W^{PV} &=& v_L R^L_{AV} +v_T R^T_{AV} +v_{T'}R^{T'}_{VA}
\end{eqnarray}
by the transverse projection of the function $X(\theta,\tau,\psi;\eta_F)$
defined in that work for the EM case, denoted $X_T$, and by its PV analog,
denoted ${\widetilde X}_T$, respectively:
\begin{eqnarray}
  X_T(\theta,\tau,\psi;\eta_F) &=& v_T U^T \\
  {\widetilde X}_T(\theta,\tau,\psi;\eta_F) &=& a_A v_T {\widetilde U}^T \ ,
\label{eq:XT}
\end{eqnarray}
where, using the notation of refs.~\cite{Alb88,Don92a}
\begin{eqnarray}
  U^T &=& 2W_1 + W_2\Delta
\label{eq:UTA} \\
  {\widetilde U}^T &=& 2{\widetilde W}_1 + {\widetilde W}_2\Delta\ .
\label{eq:UTB}
\end{eqnarray}
In the above
\begin{eqnarray}
  W_1(\tau) &=& \tau G_M^2(\tau) \\
  W_2(\tau) &=& {1\over1+\tau}\left[G_E^2(\tau)+\tau G_M^2(\tau)\right] \\
  {\widetilde W}_1(\tau) &=& \tau G_M(\tau){\widetilde G_M}(\tau) \\
  {\widetilde W}_2(\tau) &=& {1\over1+\tau}\left[
   G_E(\tau){\widetilde G}_E(\tau)+\tau G_M(\tau){\widetilde G}_M(\tau)\right],
\end{eqnarray}
so that
\begin{eqnarray}
  (1+\tau)W_2(\tau) - W_1(\tau) &=& G_E^2(\tau) \\
  (1+\tau){\widetilde W}_2(\tau) - {\widetilde W}_1(\tau) &=&
   G_E(\tau){\widetilde G}_E(\tau) \ .
\end{eqnarray}
The quantity $\Delta$, the vector current analog of (\ref{eq:Cdelta}),
was defined in ref.~\cite{Alb88}:
\begin{equation}
  \Delta = {\tau\over\kappa^2}\left[ {1\over 3}(\varepsilon_F^2 +
   \varepsilon_F\Gamma + \Gamma^2) + \lambda(\varepsilon_F + \Gamma) +
   \lambda^2 \right] -(1+\tau)\ .
\end{equation}
It is straightforward to show that
$\Delta$ and ${\widetilde \Delta}$ may be written in terms of
$\tau$, $\eta_F$ and the scaling variable $\psi$ introduced in
ref.~\cite{Alb88}:
\begin{equation}
  \psi=\left[{1\over\xi_F}(\gamma_- - 1)\right]^{1/2} \times
    \left\{
    \begin{array}{lr}
       +1, & \lambda\ge\lambda_0 \\
       -1, & \lambda\le\lambda_0
    \end{array}
    \right.
\end{equation}
where
\begin{equation}
  \gamma_- = \kappa\sqrt{1+1/\tau}-\lambda, \quad \xi_F=\epsilon_F-1,
    \quad {\rm and} \quad \lambda_0=[\sqrt{1+4\kappa^2}-1]/2 \,.
\end{equation}
Since $\Delta$ and ${\widetilde \Delta}$ are both of order $\eta_F^2<\!<1$,
excellent approximations for $X_T$ and ${\widetilde X}_T$
may be obtained by dropping the terms containing them in the expressions above.
Of course, as usual it is intended that we take one copy of these
expressions for the proton contribution (${\cal N}=Z$) and add it to
another for the neutron contribution (${\cal N}=N$).
We are thus led to consider the first of our new observables:
\begin{equation}
  {\cal R}_1(q,\theta) \equiv {
    \int_{\omega_{min}}^{\omega_{max}}\!\!d\omega\ W^{PV}(q,\omega)
    \big/\widetilde{X}_T(\theta,\tau,\psi;\eta_F) \over
    \int_{\omega_{min}}^{\omega_{max}}\!\!d\omega\ W^{EM}(q,\omega)
    \big/X_T(\theta,\tau,\psi;\eta_F)
  } \,.
\label{eq:sigmatilde}
\end{equation}

While the above ideas represent one way to proceed, other approaches can
also be taken: let us turn to a second of these before presenting detailed
results using several nuclear models.
Motivated by sum rules instead of $y$--scaling (especially by the
Coulomb Sum Rule (CSR) --- see ref.~\cite{Don89a}), let us
recall the general expression of the inclusive
cross section for inelastic electron scattering in the free RFG model:
\begin{equation}
  {d^2\sigma\over d\Omega d\epsilon'} =
    {{\cal N}\sigma_M\over4m_N\kappa} S(\psi)
    \left\{ v_L U^L + v_T U^T \right\} \ ,
\label{eq:cross-sect}
\end{equation}
where we now have the longitudinal contribution (the analog of (\ref{eq:UTA}))
\begin{equation}
  U^L = {\kappa^2\over\tau}(G_E^2 + W_2\Delta)
\end{equation}
(see ref.~\cite{Don92a}) and $X_L=v_L U^L$.
In (\ref{eq:cross-sect}) the $\psi$--scaling function is given by
\begin{equation}
  S(\psi) = {3\xi_F\over\eta_F^3}(1-\psi^2)\theta(1-\psi^2)\ .
\end{equation}
Moreover, to insure that the asymptotic value of unity for the sum rules is
reached at large $q$, we notice that
\begin{equation}
  \int_{-1}^1\!d\psi\,(1-\psi^2) = {4\over3}.
\end{equation}
Accordingly, one has
\begin{equation}
  {3\over8m_N}\int_0^\infty\!d\omega\,
    \left({\partial\psi\over\partial\lambda}\right)(1-\psi^2)
    \theta(1-\psi^2) = 1.
\end{equation}

It is thus clearly apparent that the convenient choice is to introduce the
following EM longitudinal and transverse normalized nuclear
responses
\begin{equation}
  S^{L,T} \equiv  v_{L,T} R^{L,T}\big/ X'_{L,T}\ ,
\label{eq:SLT}
\end{equation}
where the normalizing factors are given by
\begin{equation}
  X'_{L,T} \equiv {{\cal N} X_{L,T} \over(\kappa\eta_F^3/2\xi_F)
  (\partial\psi/\partial\lambda)}\ .
\label{eq:divfac}
\end{equation}
Similarly, in discussing the PV responses, following ref.~\cite{Don92a}
we need in addition to (\ref{eq:UTB}) two more functions
\begin{eqnarray}
  {\widetilde U}^L &=& {\kappa^2\over\tau}(G_E{\widetilde G}_E
   + {\widetilde W}_2\Delta)   \\
  {\widetilde U}^{T'} &=& \sqrt{\tau(1+\tau)}{\widetilde W}_3
   [1+{\widetilde\Delta}]  \ ,
\end{eqnarray}
where ${\widetilde\Delta}$ is given in (\ref{eq:Cdelta}) and
${\widetilde W}_3(\tau)=2G_M(\tau)G_A(\tau)$.  The corresponding
normalized nuclear responses are
\begin{eqnarray}
  {\widetilde S}^{L,T} &\equiv& v_{L,T}R^{L,T}_{AV}\big/
   {\widetilde X}'_{L,T} \\
  {\widetilde S}^{T'} &\equiv& v_{T'}R^{T'}_{VA}\big/
   {\widetilde X}'_{T'} \ ,
\end{eqnarray}
where, as above,  the normalizing factors are given by
\begin{eqnarray}
   {\widetilde X}'_{L,T} &\equiv& {{\cal N} {\widetilde X}_{L,T}
    \over(\kappa\eta_F^3/2\xi_F) (\partial\psi/\partial\lambda)}\\
   {\widetilde X}'_{T'} &\equiv&  {{\cal N} {\widetilde X}_{T'}
    \over(\kappa\eta_F^3/2\xi_F) (\partial\psi/\partial\lambda)}
\end{eqnarray}
with ${\widetilde X}_L=a_A v_L{\widetilde U}^L$, ${\widetilde X}_T
=a_A v_T{\widetilde U}^T$ and ${\widetilde X}_{T'}=a_V
v_{T'}{\widetilde U}^{T'}$.
The derivative in the above equations is given by
\begin{eqnarray}
  {\partial\psi\over\partial\lambda} &=& {\kappa\over\tau}
   {\sqrt{1+\xi_F\psi^2/2}\over{\sqrt{2\xi_F}}}\left[ {{1+2\lambda+\xi_F\psi^2}
   \over{1+\lambda+\xi_F\psi^2}} \right] \\
  &=& {\kappa\over{\eta_F\tau}}\left( {{1+2\lambda}\over
   {1+\lambda}} \right) + {\cal O}[\eta_F^2]\ .
\end{eqnarray}
Finally, we obtain a set of five sum rules\footnote{In fact, it is
not practical experimentally to integrate over the full range of
$\omega$; the sum rules used in the present work, as well as
ref.~\cite{Don89}, should be understood to involve integrations over
the usual quasielastic response region.  While this is well defined
in our model, there must always be some doubt as to whether this has
been achieved experimentally.  In the latter case, the longitudinal
response is reasonably confined to the region defined by the RFG model
and yet could have strength extending to high $\omega$ that is
essentially unmeasurable.  Consequently, when we give the range of
integration as extending up to infinity, we actually mean extending
to a high enough value of $\omega$ that the response function has
peaked and fallen back essentially to zero.}:
\begin{eqnarray}
  \int_0^\infty\!d\omega\, S^K(q,\omega) &=& 1,\ \ \ \ K=L,\ T \\
  \int_0^\infty\!d\omega\, {\widetilde S}^K(q,\omega) &=& 1,
   \ \ \ \ K=L,\ T,\ T',
\end{eqnarray}
for $\kappa>\eta_F\leftrightarrow q>2p_F$ and are now in a position to define
the second of our new observables:
\begin{equation}
  {\cal R}_2(q,\theta) \equiv {
    \int_{\omega_{min}}^{\omega_{max}}\!\!d\omega\ W^{PV}(q,\omega)
    \big/\widetilde{X}'_T(\theta,\tau,\psi;\eta_F) \over
    \int_{\omega_{min}}^{\omega_{max}}\!\!d\omega\ W^{EM}(q,\omega)
    \big/X'_T(\theta,\tau,\psi;\eta_F)
  } \,.
\label{eq:sigmatildx}
\end{equation}

Let us begin the discussion of our results by considering the Coulomb
sum rule.  For clarity,
retaining only the leading dependence in expansions in powers
of $\eta_F$ (although in the results to follow we use the exact
expressions), we obtain
\begin{eqnarray}
  \Sigma^L(q) &\equiv& \int_0^\infty\!d\omega\, S^L(q,\omega)
   \label{eq:CSRex} \\
   &\cong& \int_0^\infty\!d\omega\, {{R^L(q,\omega)}
   \over{\left( {{1+\lambda}\over{1+2\lambda}} \right)
   \left[Z G_{Ep}^2 + N G_{En}^2 \right] }}
   \label{eq:CSR} \\
  &\rightarrow& 1 + {\cal O}[\eta_F^2]\ \ \ \ \ \ {\rm RFG,}\ q>2p_F\ ,
\end{eqnarray}
in agreement with eq.~(3.31) of ref.~\cite{Don89a} and with
ref.~\cite{DeF84} at the QEP where $\lambda=\tau$.  Note that at
$q=2p_F=550$ MeV/c, the factor $(1+\lambda)/(1+2\lambda)\cong0.94$ and
therefore a na{\"\i}ve sum rule obtained by neglecting this factor
(as is sometimes done in analyzing experimental data) will be about
6\% too low at $2p_F$ and even further below unity at higher $q$.
In Fig.~8 we show the CSR as a function of $q$ computed using the
exact expression (\ref{eq:SLT}): as expected by the way that we have
constructed it, the free RFG answer becomes exactly unity at $q=2p_F$
and remains so for all higher values of $q$.  The CSR for our more
sophisticated model that includes pionic correlation and MEC contributions
approaches unity from below, falling about 9\% below the asymptotic
answer at $q=2p_F$.  Thus, the above relativistic effect involving
$\lambda$ and the pionic effects together yield a 15\% decrease of the
CSR from the na{\"\i}ve answer at $q=2p_F$.  Indications are that
experiment \cite{YYY} yields a result that is even smaller, although
the comparison with our pionic model is quite encouraging.  The other
curves in Fig.~8 will be discussed below.

In Fig.~9 we show the energy shifts of the peaks of the $R^L$ and $R^T$
responses away from the free RFG value:
\begin{equation}
  {\bar \epsilon}_{L,T}(q) = \omega\left[ {\rm peak\ in\ }R^{L,T} \right]
   -\omega\left[ {\rm peak\ in\ RFG} \right] \ ,
\label{eq:Shift}
\end{equation}
drawing upon the results presented in ref.~\cite{Alb93} for the model
in which pionic correlations and MEC effects have been included (again,
the other curves will be discussed below).  A $q$--dependent hardening
of the positions of the maxima of the responses is seen which, for the
reasons presented in ref.~\cite{Alb93}, is greater for $R^L$ than for
$R^T$.  As $q$ increases, this hardening becomes weaker --- again there
is some evidence from experiment \cite{Wil93} for this behaviour, which
helps to substantiate the pionic approach that we are following.

Let us now turn to a discussion of the quantities ${\cal R}_1$ and
${\cal R}_2$ defined above.  In Fig.~10 we show ${\cal R}$ (actually
${\cal R}_2$ is shown; however, since ${\cal R}_1$ and ${\cal R}_2$
differ by a negligible amount for the kinematics chosen here, we shall
consistently only show one of the two ratios) as a function of $\theta$
for three values of $q$.  As in Figs.~3, 4 and 7 two families of curves
are displayed, one for the free relativistic Fermi gas (RFG) and one
for the model with pionic effects included ($\pi$), and each family
has three curves ($g_A^{(1)}=1.26$, $1.26+10$\% and $1.26-10$\%).
Looking at Fig.~10b (and especially at the inset where the backward--angle
region is expanded) we see a significant range of angles over which
the pionic effects provide negligible modifications with respect to the
RFG results and where the (merged) curves with the three values of the
isovector/axial--vector strength can clearly be discerned. This behaviour
is similar at $q=1$ GeV/c, although not quite as nicely separated; even
so the difference between the RFG and pionic families at backward
scattering angles amounts to an effective change in $g_A^{(1)}$ of only
about 4\%.  It should also be realized that the difference between the
families of curves is not the {\em uncertainty} in the curves.  In the
light of the successes seen reflected in Figs.~8 and 9, we believe the
pionic model to represent the actual state of affairs better than the
free RFG model.  Even at low momentum transfer (Fig.~10a) there is still
a range of angles where an axial--vector determination could be
attempted, although somewhat higher values of $q$ appear to be better suited
for this (see also ref.~\cite{Don92a} where it is argued that the
statistical precision would be best around 400--500 MeV/c).

In Fig.~11 similar results are shown, except now for variations in
the magnetic strangeness single--nucleon form factor (Cf. the
results in Fig.~5). Clearly, as stated previously, PV quasielastic
electron scattering is de--sensitized to the magnetic strangeness
content for the reasons presented in detail in ref.~\cite{Don92a}.
At low--to--intermediate values of $q$ the curves computed with (dashed)
and without (solid) magnetic strangeness almost coincide; importantly,
they are close enough together that the above conclusions are not
modified. By the time $q=1$ GeV/c is reached (Fig.~11c) some separation
of the curves begins to appear and thus, in principle, there is some
degree of sensitivity to changes in $\mu_s$.  On the other hand, it is
also clear \cite{Mus92,Mus93a} that PV elastic electron scattering from
the proton as in the MIT/Bates SAMPLE experiment is more ideal for
studying $G_M^{(s)}$.

The sensitivity to electric strangeness is more pronounced, however, as
can be seen in Fig.~12. As the momentum transfer increases (and
accordingly the single--nucleon electric strangeness form factor
grows, being proportional to $\tau$ at low momentum transfer ---
see (\ref{eq:GEs})) the separation between results with different
models for $G_E^{(s)}$ becomes quite noticeable and at $q=1$ GeV/c
overwhelms the spread that occurs in going from the RFG to the pionic
correlated model.  While again alternative approaches can be taken
to determine $G_E^{(s)}$, such as PV elastic electron scattering from
$^4$He (see refs.~\cite{Mus92,Mus93a}), in this case there may be some
merit in employing forward--angle PV quasielastic electron scattering
for this purpose as well (see also ref.~\cite{Had92}).

Finally, let us come to some comparisons with a different model for
the nuclear physics content in the problem, namely one where the
responses, cross section and PV asymmetry are calculated \cite{Hor93}
using the relativistic mean--field theory (RMFT) of nuclear matter \cite{SW}.
One of the variants presented in ref.~\cite{Hor93} involves the
introduction of an effective nucleon mass $m_N^*$ in specific
places in the formalism ({\em e.g.,\/} the Dirac contributions that
involve $F_1$ and the Pauli contributions that involve $F_2$ are
affected differently) --- a value of $m_N^*\cong 0.68m_N$ is
favoured to fit the effective nuclear binding energy.  We have repeated
the calculations for this model presented in ref.~\cite{Hor93} and
applied the results to the observables discussed above. Let us start with
the Coulomb sum rule: while the pionic model discussed above yields
quite encouraging results for this quantity when compared with experiment,
the RMFT CSR\footnote{Note that here we present a relativistic CSR
following the developments presented above.  In ref.~\cite{SW} where
a CSR is also discussed a non--relativistic expansion is made, retaining
only the square of the Dirac form factor in the dividing factor. The
changes that result when terms involving the Pauli form factor are
included are quite large; these are effectively incorporated in
having used Sachs form factors, as we do in the present work (see the
discussion in ref.~\cite{Alb90}), and yield the result in Fig.~8 which
then appears to differ significantly from that in ref.~\cite{SW}.
Naturally, it is largely irrelevant what prescription one adopts as
long as the same is applied to experiment.  Since the one presented in
(\ref{eq:CSR}) is essentially the one used in analyzing experiment, we
shall continue to adopt this approach.}
is not in good agreement with experiment, as can be seen by
examining Fig.~8. Likewise in Fig.~9 we see that the RMFT model
produces a shift in the QEP that is rather dramatic and, moreover, is
is disagreement with experiment.  Thus, the RMFT results should not be
taken too seriously, but rather should be used to see how sensitive
the various observables that we have constructed are to rather extreme
modeling differences.  In Fig.~13 we show results for a few
representative cases.  We include the RFG and the pionic model results
for comparison and, in addition to using $m_N^*= 0.68m_N$ in the
RMFT, we also present results with $m_N^*= 0.8m_N$ to see the trends
as functions of the effective mass. In Fig.~13a we observe that
$\Delta{\cal A}$ could serve to differential between the various
models.  Even a rather crude measurement of this quantity should be
capable of telling the difference between the pionic correlated model
and the RMFT model with $m_N^*= 0.68m_N$, for instance. The quantity
${\overline{\cal A}}$, on the other hand, is more ambiguous as can be
seen in Fig.~13b.  As discussed in ref.~\cite{Hor93}, if the
$m_N^*= 0.68m_N$ RMFT model were viable (in which case the quantity
$\rho$ introduced in (\ref{eq:ATapp}) would be about twice as large
as in the pionic model --- in other words, there would be a much more
dramatic modification of the isoscalar responses in the former case than in
the latter), then there is more of a spread in the results for this
observable and hence more confusion in attempting to extract the
dependence on the axial--vector form factor (see above). However,
such is not the case for the quantity ${\cal R}$, as can be seen from
Fig.~13c.  There all models coalesce at large angles and the above
arguments in favour of using this observable to determine the
isovector/axial--vector single--nucleon form factor remain valid. It
remains to be seen what the situation will be for more sophisticated
models when the observables we have constructed are employed, although
it would appear likely given the above arguments that, for any model
that is reasonably capable of reproducing the known behaviour of the
parity--conserving quasielastic responses (not to mention the PV
observables such as $\Delta{\cal A}$ that can serve to limit the
model uncertainties to a very large extent), this will continue to
be an effective way to proceed.

\section{Conclusions}
\label{sec:concl}

In this paper we have applied our treatment of pionic correlations
and meson--exchange currents in quasielastic electron
scattering developed in our past work on the subject to studies of
parity--violating electron scattering.  Quasielastic PV electron
scattering was also previously investigated by us in earlier studies
within the context of the relativistic Fermi gas model and the present
work extends that treatment of the problem insofar as it incorporates
the pionic effects and introduces new observables, constructed as specific
integrals involving the differential asymmetry and electromagnetic cross
section, whose purpose is to emphasize the various physics issues of
interest while suppressing unwanted model dependences.  In particular,
on the one hand we have defined an observable ($\Delta{\cal A}$) designed
to be very sensitive to fine details in the pionic correlations, while
at the same time to be rather insensitive to variations in some of the
(as yet) unmeasured single--nucleon form factors. On the other hand,
observables denoted ${\cal R}_{1,2}$ have been constructed as weighted
integrals of the asymmetry and EM cross section, based on the concepts
of $y$--scaling and sum rules, which have the reverse properties: they
are seen to be quite unaffected by pionic correlations under favourable
circumstances and yet capable of being used to provide information on
specific single--nucleon form factors, notably the isovector/axial--vector
form factor at backward angles and the electric strangeness form factor
at forward angles.  As anticipated from previous studies, there is little
sensitivity to the magnetic strangeness form factor of the nucleon in
PV quasielastic electron scattering from nuclei.

What emerges from our investigations are several possibilities: clearly
some combination of parity--conserving (EM) quasielastic electron
scattering and of observables such as $\Delta{\cal A}$ should help in
limiting the level of nuclear model uncertainty in the problem.  For
instance, even relatively modest--precision measurements of $\Delta{\cal A}$
in the
forward direction should leave little doubt about how well the model is
doing. Concerning this observable, we have discussed
how it is affected by the longitudinal PV response:
in the free RFG model it is quite small owing to delicate cancellations
which are dictated by the standard model hadronic couplings that weight
the isoscalar and isovector components occurring in this response.
However, the pion correlates the nucleons strongly in the isoscalar
channel, while doing so only weakly in the isovector channel, and as
a consequence, it turns out that for not too high momentum transfers
the pion restores the longitudinal PV response back to a size that is
comparable
to the transverse and axial--vector PV responses and gives rise to an
anomalous $\omega$--dependence (strongly negative at low-- and strongly
positive at high--energies). Because of the opposite sign of the
longitudinal and transverse PV responses, the asymmetry then comes close
to vanishing at some frequency.
Actually, were it possible to disentangle the longitudinal
contribution to the asymmetry, then experimental evidence for an
interesting dynamical restoration of the left-right parity symmetry could be
obtained.

Having limited the model uncertainty, one will be in a position to
proceed to study the single--nucleon form factors using, for example, the
energy--averaged asymmetry (${\overline{\cal A}}$) or the quantities
${\cal R}_{1,2}$ constructed from the asymmetry --- since
${\overline{\cal A}}$, ${\cal R}_1$ and ${\cal R}_2$ all
have been constructed from integrations
across the quasielastic response region, they will have the maximal
statistical precision attainable in such experiments.  For such integrated
observables it appears that a level of precision approaching 1\% might be
feasible in future PV electron scattering experiments and hence
determinations of the axial--vector single--nucleon form factor at the
5--10\% level can be contemplated. Alternatively,
one may be most interested in the nuclear many--body responses entering
into the asymmetry, in which case quantities such as $\Delta{\cal A}$ (and to
some extent, particularly in the forward direction, ${\overline{\cal A}}$
as well) should provide a sensitive probe of these nuclear response functions.
It should be remarked that, although the level of precision expected for
$\Delta{\cal A}$ is typically a few times the 1\% that might
be reached for ${\overline{\cal A}}$, ${\cal R}_1$ or ${\cal R}_2$, even
a relatively crude measurement of $\Delta{\cal A}$ would provide
interesting information about the nuclear responses.

In discussing the other side of the problem, namely studies of the
single--nucleon form factors, we have found
the observables ${\cal R}_{1,2}$ defined in this
work to be sufficiently uncontaminated by correlation effects for
favourable kinematics that the above limiting of the nuclear model
uncertainty, while desirable, may not even be required.  These observables
have specifically been designed to integrate out the anomalous
$\omega$--dependence alluded to above and so to leave quantities that,
for the most part, only retain sensitivities to variations in some (although
not all) of the single--nucleon form factors. Indeed, in testing
these ideas with other models, in particular, with responses obtained
on the basis of relativistic mean--field theory, we have found all of
the above conclusions to work in that case as well.  It should be
remarked that this last test is a rather stringent one, since we already
know that results such as the Coulomb sum rule and the ``hardening'' of
the EM responses are not well reproduced by the na{\"\i}ve RMFT
description --- the pionic model put forward in the present work does
much better in this regard.  It remains to be seen whether the observables
advocated here serve as well for other models, although our expectation is that
they will do so.  In the final analysis we continue to be encouraged
by the fact that it appears likely that interesting information about
the isovector/axial--vector and electric strangeness form factors of the
nucleon could be obtained from quasielastic PV electron scattering
experiments even when nuclear many--body effects are taken into account.


\pagebreak

\section*{Figure captions}
\begin{itemize}
\item[Fig.~1] The PV response $R^{T'}_{VA}$ versus $\omega$ for three
 values of momentum transfer, $q=300$ (a), 500 (b) and 1000 MeV/c (c).
 The dashed curve is for the free RFG and the
 solid curve is for the pionic model discussed in the text.
\item[Fig.~2] Typical MEC diagrams involving both a photon and a $Z^0$
 for $1p$--$1h$ excitations.  The contributions where the $Z^0$ involves
 only the axial--vector current are higher order and have been neglected
 in the present work.
\item[Fig.~3] The quantity $\Delta{\cal A}$ defined in the text shown as
 a function of $\theta$ for $q=300$ (a), 500 (b) and 1000 MeV/c (c).
 The relativistic Fermi gas model results are labeled (RFG), while the
 ones with pionic effects included are labeled ($\pi$). The two families
 of curves have $g_A^{(1)}=1.26$ (solid), $1.26+10$\% (dashed) and
$1.26-10$\% (dash--dot) --- see (\ref{eq:GAvec}).
\item[Fig.~4] The $\omega$--dependence of ${\cal A}$ for different
 kinematic conditions: ($q$ [MeV/c], $\theta$ [degrees]) = (300, 10):(a),
(300, 150):(b), (500, 10):(c), (500, 150):(d), (1000, 10):(e) and
(1000, 150):(f). The labeling of the curves is otherwise as in Fig.~3.
\item[Fig.~5] As for Figs.~3 and 4, except now with two models for the
 magnetic strangeness form factor of the nucleon: (solid
 curves) no strangeness and (dashed curves) $\mu_s=-1$ in (\ref{eq:GMs}).
 Only results for the case $q=$ 500 MeV/c are displayed.
\item[Fig.~6] As for Fig.~5, except now for three models for the electric
 strangeness form factor of the nucleon: (solid curves) no strangeness,
 (dash--dot curves) ($\rho_s$, $\lambda_E^{(s)}$) = (-3, 5.6)
 and (dashed curves) ($\rho_s$, $\lambda_E^{(s)}$) = (-3, 0) in
 (\ref{eq:GEs}).
\item[Fig.~7] As for Fig.~3, except now showing the energy--averaged
 asymmetry (${\overline{\cal A}}$).  Only results for the case $q=$ 500
 MeV/c are displayed.  Panel (a) shows the entire angular range, while
 panel (b) shows only the backward--angle region in greater detail.
\item[Fig.~8] The Coulomb sum rule defined in (\ref{eq:CSRex}) for the
 free relativistic Fermi gas (RFG), the pionic model ($\pi$) and the
 relativistic mean--field theory (RMFT) with two effective masses,
 $m_N^*=0.68 m_N$ (dash--dot) and $m_N^*=0.8 m_N$ (dashed).
\item[Fig.~9] The shift of the EM longitudinal ($L$) and transverse
 ($T$) response functions from the RFG peak (see (\ref{eq:Shift}))
 for the pionic model (solid) and the relativistic  mean--field theory
 (RMFT) with two effective masses (labeled as in Fig.~8).
\item[Fig.~10] The quantity ${\cal R}$ versus $\theta$ for $q=$ 300 (a),
 500 (b) and 1000 MeV/c (c).  The curves are labeled as in Fig.~3.
\item[Fig.~11] As for Fig.~10, except now for variations of magnetic
 strangeness.  The curves are labeled as in Fig.~5.
\item[Fig.~12] As for Fig.~10, except now for variations of electric
 strangeness.  The curves are labeled as in Fig.~6.
\item[Fig.~13] As for Figs.~3, 7 and 10, except now showing results
 at $q=$ 300 MeV/c for
 the free relativistic Fermi gas (RFG), the pionic model ($\pi$) and
 the relativistic mean--field theory (RMFT) with two values for the effective
 mass,  $m_N^*=0.68 m_N$ (dash--dot) and $m_N^*=0.8 m_N$ (dashed).
\end{itemize}

\end{document}